\documentclass[cite,figures,bm]{epl}

\title{Equation of state of a superfluid Fermi gas in the BCS-BEC crossover}
\shorttitle{Equation of state in the BCS-BEC crossover}
\author{H. Hu, X.-J Liu \and P. D. Drummond}

\institute{ARC Centre of Excellence for Quantum-Atom Optics,
Department of Physics,
 University of Queensland, Brisbane, QLD 4072, Australia}


\begin{document}

\maketitle

\begin{abstract}
We present a theory for a superfluid Fermi gas near the BCS-BEC crossover,
including pairing fluctuation contributions to the free energy similar
to that considered by Nozi\`{e}res and Schmitt-Rink for the normal phase.
In the strong coupling limit, our theory is able to recover the Bogoliubov
theory of a weakly interacting Bose gas with a molecular scattering
length very close to the known exact result. We compare our results
with recent Quantum Monte Carlo simulations both for the ground state
and at finite temperature. Excellent agreement is found for all interaction
strengths where simulation results are available.
\end{abstract}

The recent experimental realization of strongly interacting Fermi
gases of $^{6}$Li and $^{40}$K atoms near a Feshbach resonance has
opened up the exciting possibility of investigating the crossover
from a Bardeen-Cooper-Schrieffer (BCS) superfluid to a Bose-Einstein
condensate (BEC) \cite{crossover,cmExp,chin,kinast,ketterle}. In
these systems, the inter-atomic interaction strength can be varied
by tuning the energy of a near-resonant molecular state with a magnetic
field.

Below resonance where the $s$-wave scattering length $a$ is positive,
stable diatomic molecules are observed to form a BEC at low temperatures.
Above resonance, with $a<0$, the molecules dissociate and form a
BCS superfluid of fermionic pairs. In the crossover region where the
scattering length $a$ is large one can access a new, strongly correlated
regime known as the unitary limit \cite{unitary}. Recent experiments
in the crossover regime have found evidence for this transition by
measuring low-lying collective modes \cite{cmExp,combescot,hu} and
heat capacity \cite{kinast,chen}.

These rapid experimental developments constitute an ideal testing
ground for theoretical studies of the BCS-BEC crossover. However,
theoretical results available in the literature are limited in the
strongly correlated unitary regime. The first systematic study of
the crossover at zero temperature was provided by Eagles and Leggett
based on BCS mean-field equations \cite{eagles,leggett}. Later, the
effects of pair fluctuations were considered by Nozi\`{e}res and Schmitt-Rink
(NSR) at temperatures above the superfluid transition \cite{nsr,randeria}.
This was recently extended to the superfluid phase by Strinati \textit{et
al.} using finite temperature Green functions \cite{strinati,strinatiCmp}.

Extensions of these approaches to take into account the bare Feshbach
molecule have also been presented \cite{molecule,griffin}, with the
conclusion that additional two-channel effects can be neglected for
broad resonances and detunings near the crossover regime \cite{diener,liuhu}.
All these studies give a qualitative description of the crossover:
but none of them are quantitatively correct in the unitary limit,
and in the BEC region. A major drawback of these theories is that
the predicted value of the molecular scattering length in the deep
BEC regime, $a_{m}=2a$, does not agree with the exact result from
the solution of the four-body problem \cite{petrov,shina,brodsky},
\textit{i.e.}, $a_{m}\simeq0.60a$. This much lower value is, however, consistent
with Quantum Monte Carlo (QMC) simulations\cite{FNQMC,carlson}.

The purpose of the present Letter is to develop a \emph{quantitatively}
reliable theory for superfluid Fermi gases in the broad resonance
or single channel limit, at all interaction strengths and low enough
temperatures. To this end, we extend the NSR analysis to the superfluid
phase on top of the BCS mean field approximation. An essential ingredient
of our theory is that the number equation, \textit{i.e.}, the relation
$n=-\partial\Omega/\partial\mu$, is satisfied for the full thermodynamic
potential $\Omega$ --- not just for the mean-field contribution to
$\Omega$. In the deep BEC limit, where molecule-molecule correlations
are important, this requirement renormalizes the mean-field molecular
scattering length of $a_{m}=2a$ to a value of $a_{m}\simeq0.57a$
\cite{ohashi}, which is very close to the exact four-body prediction
\cite{petrov}. As a consequence, our results for the equation of
state at zero temperature along the full range of the BCS-BEC crossover
are in excellent agreement with the QMC data \cite{FNQMC}. The temperature
dependent results also agree with recent path integral Monte Carlo
calculations \cite{bulgac}.

The system we consider is a uniform gas of $N$ Fermi atoms in two
hyperfine states denoted as pseudo-spins $\sigma=\uparrow,\downarrow$,
with $N_{\uparrow}=N_{\downarrow}=N/2$. To characterize the superfluid
ground state, we introduce explicitly an order parameter $\Delta$
that will be determined at the mean field level, and use the Nambu
spinor representation, in which the system is described by the Hamiltonian
${\mathcal{H}}={\mathcal{H}}_{0}+{\mathcal{V}}_{\Delta}+{\mathcal{V}}_{int}$,
with the terms, \begin{eqnarray}
{\mathcal{H}}_{0} & = & \sum_{{\mathbf{k}}}\psi_{{\mathbf{k}}}^{+}\left[\xi_{{\mathbf{k}}}{\mathbf{\sigma}}_{z}-\Delta{\mathbf{\sigma}}_{x}\right]\psi_{{\mathbf{k}}}+\sum_{{\mathbf{k}}}\xi_{{\mathbf{k}}},\label{H0}\\
{\mathcal{V}}_{\Delta} & = & \sum_{{\mathbf{k}}}\psi_{{\mathbf{k}}}^{+}\Delta{\mathbf{\sigma}}_{x}\psi_{{\mathbf{k}}},\label{Hgap}\\
{\mathcal{V}}_{int} & = & \frac{U_{0}}{2}\sum_{{\mathbf{k}k}^{\prime}{\mathbf{q}}}\left(\psi_{{\mathbf{k}+q}}^{+}\sigma_{z}\psi_{{\mathbf{k}}}\right)\left(\psi_{{\mathbf{k}}^{\prime}-{\mathbf{q}}}^{+}\sigma_{z}\psi_{{\mathbf{k}}^{\prime}}\right).\label{Hint}\end{eqnarray}
 where $\psi_{{\mathbf{k}}}^{+}=(c_{{\mathbf{k}}\uparrow}^{+},c_{-{\mathbf{k}\downarrow}})$
is the Nambu creation field operator for Fermi atoms with the kinetic
energy $\xi_{{\mathbf{k}}}=\epsilon_{{\mathbf{k}}}-\mu={\hbar^{2}}{\mathbf{k}}^{2}/2m-\mu$,
$\mu$ is the chemical potential, and ${\mathbf{\sigma}}_{x}$ and
${\mathbf{\sigma}}_{z}$ are the $2\times2$ Pauli matrices. The contact
interaction $U_{0}$ for atom-atom interactions is renormalized by
introducing the $s$-wave scattering length $a$. This is defined
by the low energy limit of the two-body scattering problem via $(4\pi \hbar^2 a/m)^{-1}=U_{0}^{-1}+$
$\sum_{{\mathbf{k}}}(2\epsilon_{{\mathbf{k}}})^{-1}$.

After taking into account anti-commutators, the {}``unperturbed''
term ${\mathcal{H}}_{0}$ can be identified as the standard BCS mean
field Hamiltonian. In variational calculations, the approximate ground
state would be an eigenstate of ${\mathcal{H}}_{0}$, with $\Delta$
varied to minimize the total energy \cite{molecule}. However, this
approach is inappropriate on the BEC side of the transition, where
molecule-molecule scattering gives rise to strong density correlations
involving four particles rather than just two. The purpose of the
present Letter is to describe an approximate perturbation expansion
which takes this into account.

A diagrammatic expansion is performed in terms of the propagator of
the {}``free'' Hamiltonian ${\mathcal{H}}_{0}$, which is given
by: \begin{equation}
{\mathbf{G}}_{0}\left({\mathbf{k}},i\omega_{m}\right)=\left[\begin{array}{cc}
{\mathcal{G}}\left({\mathbf{k}},i\omega_{m}\right) & {\mathcal{F}}\left({\mathbf{k}},i\omega_{m}\right)\\
{\mathcal{F}}\left({\mathbf{k}},i\omega_{m}\right) & -{\mathcal{G}}\left({\mathbf{k}},-i\omega_{m}\right)\end{array}\right]\,\,.\label{gf0}\end{equation}
 Here ${\mathcal{G}}({\mathbf{k}},i\omega_{m})=-[i\omega_{m}+\xi_{{\mathbf{k}}}]/[\omega_{m}^{2}+E_{{\mathbf{k}}}^{2}]$
and ${\mathcal{F}}({\mathbf{k}},i\omega_{m})=\Delta/[\omega_{m}^{2}+E_{{\mathbf{k}}}^{2}]$
are, respectively, the BCS normal and anomalous Green functions, while
$E_{{\mathbf{k}}}=[\xi_{{\mathbf{k}}}^{2}+\Delta^{2}]^{1/2}$ denotes
the single particle excitation energy and $\omega_{m}=(2m+1)\pi/(\hbar\beta)$
is the fermionic Matsubara frequency with inverse temperature $\beta=1/k_{B}T$.
The associated {}``unperturbed'' thermodynamic potential then takes
the form, $\Omega_{0}=1/\beta\sum_{{\mathbf{k}},m}$Tr$\ln{\mathbf{G}}_{0}({\mathbf{k}},i\omega_{m})+\sum_{{\mathbf{k}}}\xi_{{\mathbf{k}}}=\sum_{{\mathbf{k}}}[(\xi_{{\mathbf{k}}}-E_{{\mathbf{k}}})+2/\beta\ln f(-E_{{\mathbf{k}}})]$,
where $f(x)=1/(e^{\beta x}+1)$ is the Fermi function.

\begin{figure}
\onefigure[width=2.5cm,angle=-90]{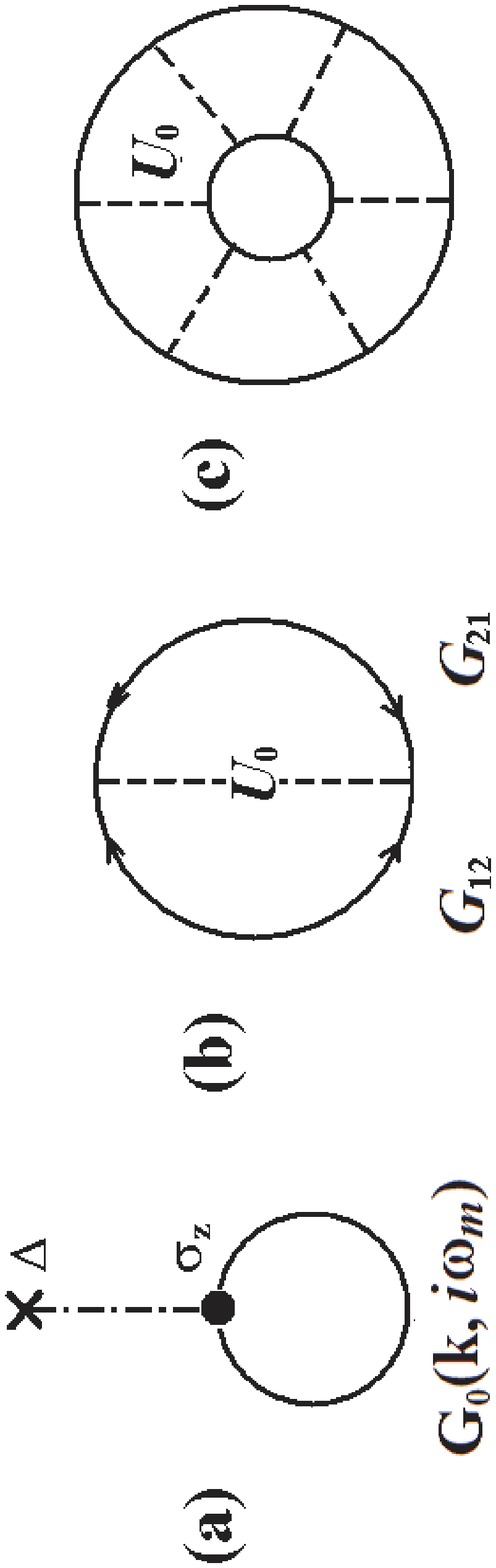}
\caption{Diagrammatic
representations of the mean field contributions {[}(a) and (b){]}
and the pairing fluctuation contributions (c) to the thermodynamic
potential. The full line represents
${\mathbf{G}}_{0}\left({\mathbf{k}},i\omega_{m}\right)$, the line
with arrows describes the anomalous Green function
$G_{12}=G_{21}={\mathcal{F}}\left({\mathbf{k}},i\omega_{m}\right)$,
and the dotted line is $U_{0}$.}
\label{Fig.1}
\end{figure}

The contributions of interactions to the thermodynamic potential consist
of a static mean field part and a fluctuation part originating from
the particle-particle Cooper channel. As shown diagrammatically in
Fig. 1a and Fig. 1b, the mean-field corrections from ${\mathcal{V}}_{\Delta}$
and ${\mathcal{V}}_{int}$ read, respectively, $\delta\Omega_{\Delta}=1/\beta\sum_{{\mathbf{k}},m}\Delta$Tr$[{\mathbf{\sigma}}_{x}{\mathbf{G}}_{0}({\mathbf{k}},i\omega_{m})]=-2\Delta^{2}/U_{0}$
and $\delta\Omega_{mf}=U_{0}[1/\beta\\\sum_{{\mathbf{k}},m}{\mathcal{F}}({\mathbf{k}},i\omega_{m})]^{2}=\Delta^{2}/U_{0}$,
where we have used the mean field gap equation:\begin{equation}
-\frac{m}{4\pi \hbar^2 a}=\sum_{{\mathbf{k}}}\left[\frac{1-2f\left(E_{{\mathbf{k}}}\right)}{2E_{{\mathbf{k}}}}-\frac{1}{2\epsilon_{{\mathbf{k}}}}\right]\,\,.\label{gapEQ}\end{equation}
 Here the coupling $U_{0}$ is eliminated in favor of the $s$-wave
scattering length $a$. These corrections, together with $\Omega_{0}$,
give rise to an overall mean field thermodynamic potential of $\Omega_{mf}=\Omega_{0}+\delta\Omega_{\Delta}+\delta\Omega_{mf}$,
\begin{equation}
\Omega_{mf}=\sum_{{\mathbf{k}}}[\xi_{{\mathbf{k}}}-E_{{\mathbf{k}}}+\frac{\Delta^{2}}{2\epsilon_{{\mathbf{k}}}}+\frac{2}{\beta}\ln f(-E_{{\mathbf{k}}})]-\frac{m\Delta^{2}}{4\pi \hbar^2 a}\,\,.\label{omegaMF}\end{equation}
 Fluctuation corrections beyond mean field are illustrated
in Fig. 1c. A sum of the resulting geometrical series thus leads to
\cite{nsr,randeria,griffin}, \begin{equation}
\Omega_{pf}=-\frac{1}{\pi}\sum_{{\mathbf{q}}}\int_{-\infty}^{+\infty}d\omega\frac{1}{e^{\beta\omega}-1}\delta\left({\mathbf{q}},\omega\right),\label{omegaPF}\end{equation}
 where, following NSR, we have written $\Omega_{pf}$ in terms of
a phase shift defined by $\delta({\mathbf{q}},\omega)=-\mathop{\textrm{Im}}\ln[-\chi_{11}({\mathbf{q}},\omega+i\eta)]-1/2\mathop{\textrm{Im}}\ln\{1-\chi_{12}^{2}({\mathbf{q}},\omega+i\eta)/[\chi_{11}({\mathbf{q}},\omega+i\eta)\chi_{11}^{*}({\mathbf{q}},-\omega+i\eta)]\}$.
Here the analytic continuation is performed and $\eta$ is a positive
infinitesimal, while $\chi_{11}=1/U_{0}+1/\beta\sum_{{\mathbf{k}},m}{\mathcal{G(}}{\mathbf{q}-k},i\nu_{n}-i\omega_{m}){\mathcal{G(}}{\mathbf{k}},i\omega_{m})$
and $\chi_{12}=1/\beta\sum_{{\mathbf{k}},m}{\mathcal{F(}}{\mathbf{q}-k},i\nu_{n}-i\omega_{m}){\mathcal{F(}}{\mathbf{k}},i\omega_{m})$
are the diagonal and off-diagonal parts of the Cooper pair propagator,
with $\nu_{n}=2n\pi/\beta$ being the bosonic Matsubara frequency.

Putting together the mean field and the pairing fluctuation corrections
to the thermodynamic potential, we obtain the total contributions,
$\Omega=\Omega_{mf}+\Omega_{pf}$. We emphasize that the gap (\ref{gapEQ})
is chosen at the mean field level, \textit{i.e.}, $\partial\Omega_{mf}/\partial\Delta=0$. Non-trivial effects beyond the
BCS mean field approximation enter into the theory through the modified
number equation $N=-\left(\partial\Omega/\partial\mu\right)_{T}$.
Explicitly, we obtain $N=N_{mf}+N_{pf,\mu}+N_{pf,\Delta}$, where
$N_{mf}=-(\partial\Omega_{mf}/\partial\mu)_{T\Delta}$, $N_{pf,\mu}=-(\partial\Omega_{pf}/\partial\mu)_{T\Delta}$,
and $N_{pf,\Delta}=-(\partial\Omega_{pf}/\partial\Delta)_{T\mu}(\partial\Delta/\partial\mu)$.
The coupled gap and particle number equations, together with the thermodynamic
potentials (\ref{omegaMF}) and (\ref{omegaPF}), form the basis of
our NSR theory in the broken-symmetry state. Once $\mu$ and $\Delta$
are obtained as a function of the interaction strength and temperature,
the entropy and the energy of the gas can then be calculated straightforwardly,
using $S=-\left(\partial\Omega/\partial T\right)_{\mu}$ and $E=\Omega+TS+\mu N$.

There is a key difference between our method and the diagrammatic
theory proposed by Strinati \textit{et al.} \cite{strinati,strinatiCmp}.
The term $N_{pf,\Delta}$, which ensures number conservation, is not
included in the diagrams considered by Strinati \textit{et al.} This
term becomes increasingly important in the BEC regime. Physically,
it generates the dominant part of the four-fermion correlations, which
are increasingly important for tightly bound Cooper pairs. This constitutes
a major advantage of our NSR theory.

To give more insight, in the strong coupling limit we re-interpret our
formalism in the framework of the functional integral method \cite{randeria}, in
which a Cooper-pair Bose field $\Delta (x,\tau )$ is introduced through the
Hubbard-Stratonovich transformation. Integrating out the fermionic degrees
of freedom in the usual fashion and setting $\Delta (x,\tau )$ $=\Delta
+\delta \Delta (x,\tau )$, the resulting effective bosonic action is then
expanded up to quadratic order in fluctuations $\delta \Delta (x,\tau )$: 
$S_{eff}\approx S^{(0)}+S^{(2)}$. After performing a Fourier transformation
we find that $S^{(0)}=\beta \Delta ^2/U_0-$Tr$\ln [-\mathbf{G}_0^{-1}]+\beta
\sum_{\mathbf{k}}\xi _{\mathbf{k}}$, and 
\begin{equation} S^{(2)}=\sum_{{\mathbf{q}}%
,n}[-\delta \Delta _q^{+}\chi _{11}({\mathbf{q}},i\nu _n)\delta \Delta
_q+\chi _{12}({\mathbf{q}},i\nu _n)(\delta \Delta _q^{+}\delta
\Delta _{-q}^{+}+\delta \Delta _{-q}\delta \Delta _q)/2]\, ,
\end{equation}
where $\delta\Delta _q=\delta \Delta ({\mathbf{q}},i\nu _n)$ is the Fourier
transformation of $\delta \Delta (x,\tau )$.  The saddle point solution of 
$S^{(0)}$gives the standard mean-field theory, while the next order
Gaussian expansion $S^{(2)}$ leads to exactly the same contribution as in
Eq. (\ref{omegaPF}). It is easy to see that in the long-wavelength and
low-frequency limits, $-\chi _{11}({\mathbf{q}},i\nu _n)\sim -i\nu _n+\hbar
^2{\mathbf{q}}^2/4m+\Delta ^2/\left( -4\mu \right) $ and $\chi _{12}({%
\mathbf{q}},i\nu _n)\sim \Delta ^2/\left( -4\mu \right) $. Hence, $S^{(2)}$
acquires the familiar form of Bogoliubov excitations \cite{keeling}. 
Our formalism therefore incorporates interactions between condensed and 
non-condensed Cooper-pairs at the level of Bogoliubov theory, which 
\emph{must} be the dominant part of four-fermion correlations at low temperature.

The crucial observation of the present Letter is that in the deep BEC limit 
($\Delta / (-\mu) \rightarrow 0$) $N_{pf,\Delta }\sim \Delta ^2/(-\mu )^{1/2}$ is of the \emph{same} order of $%
N_{mf}$, or more precisely, $N_{pf,\Delta }=CN_{mf}$ with $C\simeq 2.5$, as
shown analytically in a forthcoming publication \cite{LHD}. In contrast, 
$N_{pf,\mu }\sim \Delta ^3/(-\mu )^{3/2}\sim \Delta /(-\mu )N_{mf}$ becomes
negligible. We thereby find
that the molecular condensate density $N_B^0\simeq N/2\simeq (1+C)N_{mf}/2$.
Mean-field number and gap equations provide the expressions, $N_{mf}\simeq
\Delta ^2m^2a/(4\pi \hbar ^2)$, and $\mu _B=2\mu +\hbar ^2/(ma^2)\simeq
\Delta ^2ma^2/(2\hbar ^2)$, where $\mu _B$ is the molecular chemical
potential. Assembling together these three expressions and eliminating $%
\Delta $, one finds that $\mu _B=4\pi \hbar ^2\left[
2a/(1+C)\right] /\left( 2m\right) \times N_B^0$, implying $%
a_m=2a/(1+C)\simeq 0.57a$ --- a value close to the exact result ($a_m\simeq
0.60a$). The residual $5\%$ difference occurs because we exclude
interactions between non-condensed Cooper-pairs.

Our diagrammatic procedure is also different from the works in
Refs. \cite{petrov,brodsky}, where molecule-molecule scattering
processes are considered in isolation. These calculations perturb
around the free-fermion state and therefore require additional
terms beyond the ladder structures in Fig. 1 to obtain correct
results. By comparison, we perturb around a BCS state of
\emph{correlated} fermions \emph{below} $T_{c}$. A dominant part
of the molecule-molecule scattering is therefore already included
in the ladder diagrams due to the inclusion of $N_{pf,\Delta}$, as
we have shown  using the functional integral method. To
contain the full four-body correlations, a more complicated wave
function has to be implemented \cite{shina}. Equivalently, within
the functional integral method, we may include four-fermion
correlations by expanding the action to fourth order of the 
gap \cite{ohashi}. However, this calculation is difficult to
extend to the crossover regime.
\begin{figure}
\onefigure [width=8cm]{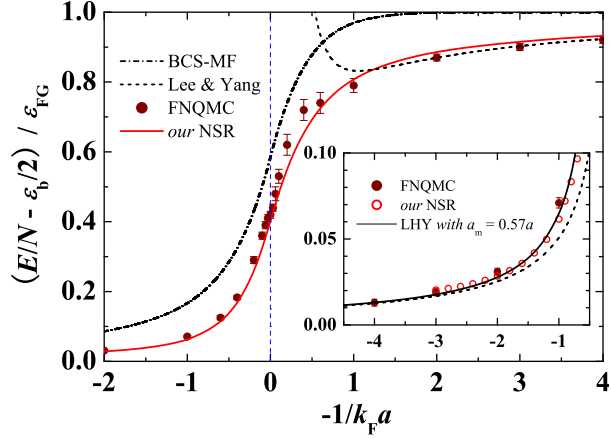} \caption{(color online) Energy
per particle $E/N$, in the BCS-BEC crossover (with the binding
energy $\epsilon_{b}=-\hbar^{2}/ma^{2}$ subtracted). Lines plotted
are: our result (solid line); fixed node QMC data from Ref.
\cite{FNQMC} (circles with error bars), Lee and Yang perturbation
theory (dashed line); BCS mean field predictions (dot-dashed
line). Inset: enlarged view of the BEC regime. The solid line
corresponds to the equation of state of a repulsive gas of
molecules given by LHY \cite{LHY},
$(E/N-\varepsilon_{b}/2)/\epsilon_{FG}=5/(18\pi)k_{F}a_{m}[1+128/(15\sqrt{6\pi^{3}})(k_{F}a_{m})^{3/2}+...]$,
with our analytic result of $a_{m}=0.57a$.}
\label{Fig.2}
\end{figure}

\begin{figure}
\onefigure[width=8cm]{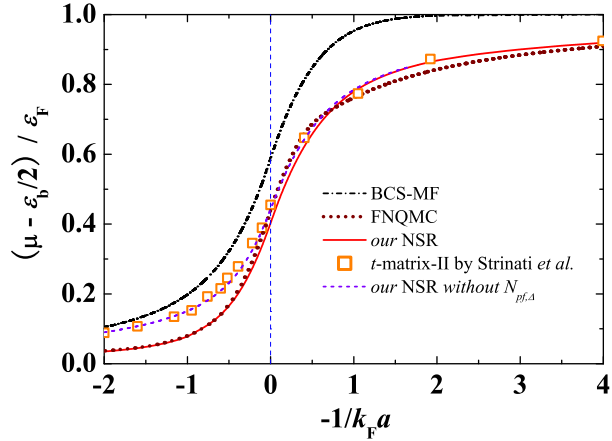} 
\caption{(color online) Chemical potential in the BCS-BEC crossover, as
predicted by various approaches. The QMC curve in dotted line is calculated
from a best fit to the QMC energies, as outlined in Ref. \cite{FNQMC}. The
diagrammatic predictions (empty squares) by Strinati {\it et al.} are taken
from Ref. \cite{strinatiCmp} without the inclusion of the self-energy shift.}
\label{Fig3}
\end{figure}

Figure 2 presents our results for the energy per particle as a function
of the interaction strength $1/k_{F}a$ at a low temperature $T=0.02T_{F}$.
The energy scale is given by the noninteracting energy, $\epsilon_{FG}=(3/5)\epsilon_{F}=(3/10)\hbar^2 k_{F}^{2}/m$,
where $k_{F}=(3\pi^{2}n)^{1/3}$ is the Fermi wave vector. As a benchmark,
the approximate fixed node QMC data at zero temperature is shown \cite{FNQMC},
together with the perturbation results $E/(N\epsilon_{FG})=1+10/(9\pi)k_{F}a+4/(21\pi^{2})(11-2\ln2)(k_{F}a)^{2}$
obtained by Lee and Yang \cite{leeyang}. In the BCS region, $1/k_{F}a<-0.5$,
we find that our results agree with the QMC data, apart from residual
differences due to the energy dependence of the scattering amplitude.
In the BEC region, $1/k_{F}a>0.5$, our predictions coincide perfectly
with QMC calculations, and thereby agree with the equation of state
of a repulsive gas of molecules with $a_{m}\simeq0.60a$, as shown
in the inset. In the range $-0.5\stackrel{\textstyle<}{\sim}1/(k_{F}a)\stackrel{\textstyle<}{\sim}0.5$ spanning
the most interesting crossover region, our results differ only slightly
from that of QMC simulations. In particular, in the unitary limit
we predict $E/N=\xi\epsilon_{FG}$ with $\xi=0.401$, compatible with
the QMC findings $\xi=0.42(1)$ \cite{FNQMC}. The overall agreement
between the two alternative calculations is therefore excellent, especially
in the challenging strong coupling regime.

We have also calculated the chemical potential at the BCS-BEC crossover,
predicted by different theories (Fig. 3). We find again an excellent quantitative
agreement between our NSR results and QMC calculations. To emphasize
the similarity between our formalism and the diagrammatic theory given
by Strinati \textit{et al.} \cite{strinati}, we compared the results
obtained \emph{without} the inclusion of the crucial term $N_{pf,\Delta}$
in the number equation, with Strinati's diagrammatic findings \cite{strinatiCmp}.
These asymptotically approach the mean field predictions in the deep
BEC limit, which would (incorrectly) imply that $a_{m}=2a$. This
observation thereby unambiguously verifies that $N_{pf,\Delta}$ is
responsible for obtaining the correct molecular scattering length.

\begin{figure}
\onefigure[width=8cm]{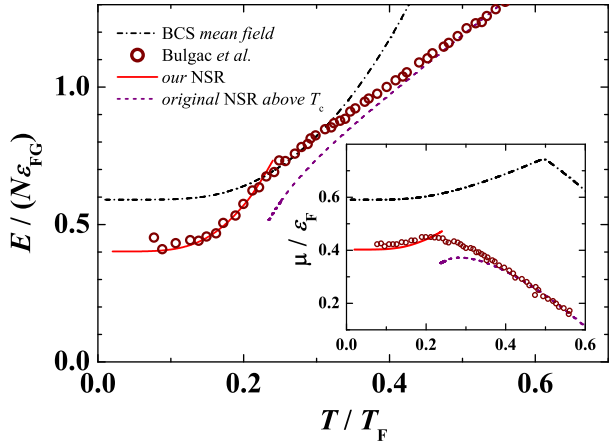} 
\caption{(color online)
Temperature dependence of the energy per particle (main plot) and
the chemical potential (inset) in the unitary limit. Lines plotted
are: our results (solid lines); path integral Monte Carlo data
from Ref. \cite{bulgac} (circles); BCS mean field (dot-dashed
lines); the original NSR prediction in Ref. \cite{nsr,randeria}.}
\label{Fig4}
\end{figure}

Finally, in Fig. 4, we compare our predictions at finite temperature
with recent path integral Monte Carlo calculations of spin 1/2 fermions \cite{bulgac}
in the strongly coupled unitary limit. At low temperature up to $T_{c}\approx0.22T_{F}$,
these results are in good qualitative agreement with each other. The
residual discrepancy is possibly due to finite size effects in
the  simulations.

To conclude, we have presented an NSR-type formalism for a Fermi gas
at the BCS-BEC crossover, in the broken-symmetry phase. A notable
achievement of our formalism is that a Bogoliubov theory of composite
Cooper pairs is reproduced in the BEC limit, with a molecular scattering
length very close to the exact value. We have compared our predictions
of the equation of state of the gas with available Monte Carlo calculations,
and find excellent agreement. Our results also make quantitative contact
with a previous diagrammatic theory in the weak and intermediate coupling
regimes. We believe, therefore, that the present formalism provide
a quantitatively reliable description of superfluid Fermi gases at
low temperature over the entire range of the crossover.

\acknowledgments
We gratefully acknowledge Prof. S. Giorgini and
Dr. P. Pieri for sending us the data file of the figures in their
references, and Dr. Shina Tan for useful discussions. We also
thank Prof. A. Griffin for bring our attention to Ref.
\cite{keeling}. Funding for this research was provided by an
Australian Research Council Center of Excellence grant.

\end{document}